\documentclass[prl,aps,twocolumn,superscriptaddress,showpacs]{revtex4-1}
\usepackage{amsmath}
\usepackage{bm}
\usepackage{graphicx}
\usepackage{color}
\usepackage{hyperref}

\newcommand{\mi}{\mathrm{i}}

\begin{document}

\title{Exciton-photon correlations in bosonic condensates of exciton-polaritons}
\author{A. V. Kavokin}
\affiliation{Russian Quantum Center, Novaya 100, 143025 Skolkovo, Moscow Region, Russia}
\affiliation{Spin Optics Laboratory, St.-Petersburg State University, 198504, Peterhof,
St.-Petersburg, Russia}
\affiliation{School of Physics and Astronomy, University of Southampton, SO17 1NJ
Southampton, United Kingdom}
\author{A. S. Sheremet}
\affiliation{Russian Quantum Center, Novaya 100, 143025 Skolkovo, Moscow Region, Russia}
\affiliation{Department of Theoretical Physics, St-Petersburg State Polytechnic
University, 195251, St.-Petersburg, Russia}
\author{I. A. Shelykh}
\affiliation{Science Institute, University of Iceland, Dunhagi-3, IS-107, Reykjavik,
Iceland}
\affiliation{Division of Physics and Applied Physics, Nanyang Technological University,
637371, Singapore}
\author{P. G. Lagoudakis}
\affiliation{School of Physics and Astronomy, University of Southampton, SO17 1NJ
Southampton, United Kingdom}
\author{Y. G. Rubo}
\affiliation{Instituto de Energ\'{\i}as Renovables, Universidad Nacional Aut\'onoma de
M\'exico, Temixco, Morelos, 62580, Mexico}

\date{December 22, 2014}

\begin{abstract}
Exciton-polaritons are mixed light-matter quasiparticles. We have developed a
statistical model describing stochastic exciton-photon transitions within a condensate
of exciton polaritons. We show that the exciton-photon correlator depends on the
``hidden variable'' which characterizes the rate of exciton-photon transformations in
the condensate. We discuss implications of this effect for the quantum statistics of
photons emitted by polariton lasers.
\end{abstract}

\maketitle

\emph{Introduction.---}A pulse of light entering a crystal in the vicinity of the exciton
resonance excites combined light-matter waves, as it was discovered by Pekar in 1957
\cite{Pekar57}. Quantization of these waves yields mixed light-matter quasiparticles,
currently known as exciton-polaritons. A classical theory developed by Hopfield in 1960s
describes this process characterizing crystals by modified dielectric functions having
resonant features \cite{Hopfield}. The quantum theory developed by Agranovich
\cite{Agranovich} considers a polariton as a chain of virtual emission-absorption acts: a
photon is absorbed creating an exciton, then an exciton recombines emitting a photon,
etc. It is usually assumed that neither emission nor absorption of photons by excitons
does take place in reality: photons entering a crystal are immediately transformed into
exciton-photon quasiparticles which may be detected as excitons or as photons with
certain probabilities given by so-called Hopfield coefficients. It is impossible to know
a priori if an exciton-polariton would collapse to exciton or a photon state in each
particular measurement.

The interpretation of exciton-polaritons as superposition light-matter quantum states is
essential for understanding of the phenomenon of polariton lasing \cite{Imamoglu}. In
polariton lasers, stimulated scattering of exciton-polaritons in semiconductor
microcavities \cite{Microcavities} leads to macroscopic population of a single quantum
state, which forms a condensate (or Bose-Einstein condensate\cite{DengRMP,Butov}) of
exciton-polaritons. A polariton condensate spontaneously emits the monochromatic and
coherent light, which constitutes the polariton lasing effect \cite{Imamoglu,Porras2002}.
Polariton lasers have been experimentally realized in various semiconductor systems
\cite{KasprzakNature,BaliliScience,Bajoni2008,Schneider2013} and operate up to room
temperature\cite{Christopolous,Lidzey1999}. From the point of view of the quantum
theory\cite{Fabrice}, spontaneous emission of each individual photon does not collapse
the many-body wave-function of the polariton condensate, which opens room for studying
the internal structure of this mixed-light matter many-body state by measuring the
statistics of emitted photons \cite{Deng2002,Schwendimann2008,Horikiri2010}.

Here we present the model of exciton-polariton system describing both coherent and
stochastic transformations of excitons and photons. Coherent processes are characterized
by the Rabi frequency $\Omega$, i.e., the splitting of upper and lower exciton-polariton
eigenmodes, while the acts of stochastic conversion happen with a characteristic time
$\tau_{xc}$. These stochastic conversions are necessarily present in the weak
exciton-light coupling regime. We assume, that they may also be present in some extent in
the strong coupling regime, in which case the exciton-photon conversion rate constitutes
a kind of ``hidden parameter'' of the system. In order to reveal the role of this hidden
parameter in the noise spectra of photoluminescence and photocurrent generated by
exciton-polaritons, one can place a condensate of exciton polaritons in a biased
semiconductor microcavity structure\cite{Savvidis}, where polaritons can decay as
excitons by electron and hole tunneling through the barriers or as photons by photon
tunneling through dielectric mirrors of the microcavity thus contributing to the
photoluminescence signal. We quantitatively analyze the correlations between the noises
in exciton and photon decay channels as functions of the stochastic exciton-photon and
photon-exciton conversion rate $\tau_{xc}^{-1}$. We discuss the potential impact of the
stochastic processes on the second order coherence in the emission of polariton lasers
and compare the results of our statistical model with the experimental data by Kasprzak
\emph{et al.} \cite{Richard}.

\emph{Basic equations.---}The strong exciton-photon coupling regime in microcavities
manifests itself in the appearance of a Rabi doublet composed by lower (LP) and upper
polariton (UP) branches. The lower $\left|LP\right>$ and upper $\left|UP\right>$
single-polariton states can be represented as
\begin{equation}\label{doublet}
\left\vert LP\right\rangle =C_{x}\left\vert X\right\rangle -C_{c}\left\vert
C\right\rangle ,\quad \left\vert UP\right\rangle =C_{c}\left\vert
X\right\rangle +C_{x}\left\vert C\right\rangle ,
\end{equation}%
where $\left|X\right>$ and $\left|C\right>$ denote the exciton and cavity quantum states,
respectively. The Hopfield coefficients are given by $C_{x,c}=2^{-1/2}(1\pm {\Delta
}/\sqrt{4\Omega ^{2}+\Delta ^{2}})^{1/2}$ with $\Delta $ being the detuning between the
bare photon and exciton mode \cite{Microcavities}. In the following we shall restrict
ourselves to the simplest $\Delta=0$ case.

The system of exciton-polaritons is described by the density matrix $\hat{%
\rho}$, which evolves according to the equation
\begin{multline}\label{rho}
 \frac{d\hat{\rho}(t)}{dt}=\frac{i}{\hbar}[\hat{\rho},\hat{H}_{0}] \\
 -\frac{1}{2}\sum\limits_{j}\frac{1}{\tau_{j}}\left(
 \hat{A}_{j}^{\dag}\hat{A}_{j}\hat{\rho} + \hat{\rho}\hat{A}_{j}^{\dag}\hat{A}_{j}
 - 2 \hat{A}_{j}\hat{\rho}\hat{A}_{j}^{\dag}
 \right).
\end{multline}
Here we assume that exciton-polaritons are excited at $t=0$ and all decay processes are
present only at $t>0$. The decay processes under consideration are described by four
Lindblad terms with $j=x,c,xc,cx$. Denoting by $\hat{a}^{\dag}$ ($\hat{a}$) and
$\hat{b}^{\dag}$ ($\hat{b}$) the creation (annihilation) operators for excitons and
photons, respectively, we describe the escape (annihilation) of an exciton with the
operator $\hat{A}_{x}=\hat{a}$ and the characteristic life-time $\tau _{x}$, the escape
of a photon is described with $\hat{A}_{c}=\hat{b}$ and the life-time $\tau _{c}$. The
two processes of conversion are present: the exciton-to-photon conversion with
$\hat{A}_{xc}=\hat{b}^{\dag }\hat{a}$ and the photon-to-exciton conversion with
$\hat{A}_{cx}=\hat{a}^{\dag }\hat{b}$, both having the same characteristic conversion
time $\tau _{xc}$. The equation for the density matrix is completed by the term
describing the coherent exciton-photon coupling with
\begin{equation}\label{Ham0}
 \hat{H}_{0}=\Omega( \hat{a}^{\dag}\hat{b} + \hat{b}^{\dag}\hat{a} ).
\end{equation}

Stochastic exciton-photon conversion makes evolution of $\hat{\rho}$ to be quite complex,
as it can be seen from the Glauber-Sudarshan representation of Eq.~\eqref{rho}. Expanding
of $\hat{\rho}$ on the basis of coherent states $\left|\mathbf{x},\mathbf{y}\right>$,
where $\mathbf{x}$ and $\mathbf{y}$ are 2D vectors such that
 $\hat{a}\left|\mathbf{x},\mathbf{y}\right>
 =(x_1+\mi x_2)\left|\mathbf{x},\mathbf{y}\right>$ and
 $\hat{b}\left|\mathbf{x},\mathbf{y}\right>
 =(y_1+\mi y_2)\left|\mathbf{x},\mathbf{y}\right>$,
one can obtain
\begin{equation}\label{GSrepres}
 \hat{\rho}=\int d^2x\,d^2y \, p(\mathbf{x},\mathbf{y},t)
 \left|\mathbf{x},\mathbf{y}\right>\left<\mathbf{x},\mathbf{y}\right|,
\end{equation}
\begin{multline}\label{FokPlanck}
  \frac{\partial p}{\partial t}
 =\frac{1}{2\tau_x}(\bm{\nabla}_x\cdot\mathbf{x})p
 +\frac{1}{2\tau_c}(\bm{\nabla}_y\cdot\mathbf{y})p
 \\
 +\Omega\hat{\mathcal{C}}p +\frac{1}{4\tau_{xc}}(\hat{\mathcal{C}}^2+\hat{\mathcal{D}}^2)p.
\end{multline}
Here operator
\begin{equation}\label{OperC}
 \hat{\mathcal{C}}=[\mathbf{x}\times\bm{\nabla}_y]+[\mathbf{y}\times\bm{\nabla}_x]
\end{equation}
describes the Rabi oscillation in the system, while
\begin{equation}\label{OperD}
 \hat{\mathcal{D}}=(\mathbf{x}\cdot\bm{\nabla}_y)-(\mathbf{y}\cdot\bm{\nabla}_x)
\end{equation}
governs the stochastic mixing between exciton and photon subsystems.

The effect of conversion can be analyzed in the limiting case
$\tau_{x},\tau_{c}\gg\tau_{xc}$, where the first two terms in the right-hand-side of
\eqref{FokPlanck} can be omitted. The remaining Fokker-Planck equation conserves the
average number of particles
 $N=\int d^{2}x\,d^{2}y\,(\mathbf{x}^{2}+\mathbf{y}^{2})p(\mathbf{x},\mathbf{y},t)$,
having in mind that
 $\hat{\mathcal{C}}\mathbf{x}^2+\hat{\mathcal{C}}\mathbf{y}^2
 =\hat{\mathcal{D}}\mathbf{x}^2+\hat{\mathcal{D}}\mathbf{y}^2=0$.
Moreover, any function $p(\mathbf{x}^{2}+\mathbf{y}^{2})$ represents a time-independent
solution. The system evolves to reach the state that corresponds to the equidistribution
of exciton and photons, what also implies the equidistribution between upper and lower
polariton branches.

The operators $\hat{\mathcal{C}}$ and $\hat{\mathcal{D}}$ obey the following useful
relations:
 $\hat{\mathcal{C}}(\mathbf{x}\cdot\mathbf{y})=0$,
 $\hat{\mathcal{D}}^2(\mathbf{x}\cdot\mathbf{y})=-4(\mathbf{x}\cdot\mathbf{y})$.
These relations can be used to obtain the time-dependence of the total energy of the
system
 $E(t)=2\Omega\int d^2x\,d^2y \, (\mathbf{x}\cdot\mathbf{y}) p(\mathbf{x},\mathbf{y},t)$,
which decays exponentially to zero, which corresponds to the bare exciton (photon) energy
in our notation, $E(t)\propto\exp(-t/\tau_{E})$, with
\begin{equation}\label{tauE}
 \frac{1}{\tau_E}
 =\frac{1}{2}\left(\frac{1}{\tau_x}+\frac{1}{\tau_c}\right)+\frac{1}{\tau_{xc}}.
\end{equation}
We note that for finite $\tau_{x,c}$ the average number of particles also decays
exponentially, $N(t)\propto\exp(-t/\tau_{N})$, but on a different time scale
\begin{equation}\label{tauN}
 \frac{1}{\tau_{N}}
 =\frac{1}{2}\left(\frac{1}{\tau_{x}}+\frac{1}{\tau_{c}}\right).
\end{equation}
As a result, the stochastic conversion strongly affects the energy per particle $E/N$,
which is otherwise time-independent for $\tau _{xc}=0$ [see the insets in
Fig.~\ref{fig2}(a) below].

In what follows, several correlators of interest will be defined by the diagonal elements
of the density matrix $P(n_{a},n_{b},t)$ in the basis of exciton and photon number
states, which describe the probabilities to have $n_{a}$ excitons and $n_{b}$ photons in
the system. The dynamics of $P(n_{a},n_{b},t)$ is found from \eqref{rho} to be
\begin{eqnarray}\label{probability}
 &&\frac{dP(n_{a},n_{b})}{dt}=\tau_{x}^{-1}[(n_{a}+1)P(n_{a}+1,n_{b})-n_{a}P(n_{a},n_{b})]
 \nonumber \\
 &&+\tau _{c}^{-1}[(n_{b}+1)P(n_{a},n_{b}+1)-n_{b}P(n_{a},n_{b})] \nonumber \\
 &&+\tau_{xc}^{-1}[(n_{a}+1)n_{b}P(n_{a}+1,n_{b}-1)-n_{a}(n_{b}+1)P(n_{a},n_{b})]
 \nonumber \\
 &&+\tau_{xc}^{-1}[(n_{b}+1)n_{a}P(n_{a}-1,n_{b}+1)-(n_{a}+1)n_{b}P(n_{a},n_{b})],
 \nonumber \\ && \phantom{+}
\end{eqnarray}
where the sum of its diagonal elements is normalized to unity
$\sum_{n_{a},n_{b}}P(n_{a},n_{b},t)=1$.

\emph{Exciton-photon correlations.---}The quantities of interest to be
analyzed in the remaining part of this Letter include the exciton-photon
correlator
\begin{equation}
g_{xc}(t)=\frac{\left\langle n_{a}(t)n_{b}(t)\right\rangle }{\left\langle
n_{a}(t)\right\rangle \left\langle n_{b}(t)\right\rangle },  \label{gsmallxc}
\end{equation}%
the exciton and photon second-order coherence
\begin{equation}
g_{2x}(t)=\frac{\left\langle n_{a}(t)^{2}-n_{a}(t)\right\rangle }{%
\left\langle n_{a}(t)\right\rangle ^{2}},\quad g_{2c}(t)=\frac{\left\langle
n_{b}(t)^{2}-n_{b}(t)\right\rangle }{\left\langle n_{b}(t)\right\rangle ^{2}}
\label{g2XandC}
\end{equation}%
and the generalized exciton-photon correlator expressed through the previous
three correlators
\begin{equation}
G_{xc}(t)=\frac{g_{xc}^{2}}{g_{2x}g_{2c}}=\frac{\left\langle
n_{a}(t)n_{b}(t)\right\rangle ^{2}}{\left\langle
n_{a}(t)^{2}-n_{a}(t)\right\rangle \left\langle
n_{b}(t)^{2}-n_{b}(t)\right\rangle }.  \label{Gbigxc}
\end{equation}%
The latter correlator is convenient since $G_{xc}=1$ for a polariton
condensate formed at the low-polariton branch independently of the
statistics of particles within the condensate. To prove this statement, we
express the exciton and photon operators through the polaritons operators
corresponding to the lower and the upper branches, $\hat{c}_{L}=(\hat{a}-%
\hat{b})/\sqrt{2}$ and $\hat{c}_{U}=(\hat{a}+\hat{b})/\sqrt{2}$,
respectively. Assuming that the upper polariton branch is empty, we obtain
\begin{equation}
G_{xc}=\frac{\langle \hat{a}^{\dag }\hat{b}^{\dag }ab\rangle ^{2}}{\langle
\hat{a}^{\dag 2}\hat{a}^{2}\rangle \langle \hat{b}^{\dag 2}\hat{b}%
^{2}\rangle }=\frac{\langle \hat{c}_{L}^{\dag 2}\hat{c}_{L}^{2}\rangle ^{2}}{%
\langle c_{L}^{\dag 2}c_{L}^{2}\rangle \langle \hat{c}_{L}^{\dag 2}\hat{c}%
_{L}^{2}\rangle }=1.  \label{unity}
\end{equation}

As we discussed above, the stochastic exciton-photon conversion leads to the
mixing of lower and upper polariton branches, and this is indicated by
deviation of $G_{xc}$ from unity. We consider coherent and Fock states of
polaritons at the lower polariton branch as initial states of our system.
The initial diagonal elements of the density matrix for the coherent state
can be described by the product of two Poisson distributions
\begin{equation}
P(n_{a},n_{b},0)=\frac{e^{-N}}{n_{a}!\,n_{b}!}\left( \frac{N}{2}\right)
^{n_{a}+n_{b}},  \label{PoissonDistr}
\end{equation}%
while for the Fock state they are given by
\begin{equation}
P(n_{a},n_{b},0)=\frac{1}{2^{N}}\frac{N!}{n_{a}!n_{b}!},  \label{binomDistr}
\end{equation}%
where $N$ is the average number of the lower-branch polaritons at $t=0$.

Figure \ref{fig1} shows the time evolution of the correlators $G_{xc}(t)$, $%
g_{xc}(t)$, $g_{2x}(t)$, and $g_{2c}(t)$, calculated assuming infinite
exciton and photon lifetimes. As expected for the initial coherent state, $%
g_{xc}(0)=g_{2x}(0)=g_{2c}(0)=G_{xc}(0)=1$. The subsequent change of all
considered correlators is substantial and experimentally verifiable. On the
time scale of $\tau _{xc}$ the correlators approach their saturation values,
\begin{subequations}
\label{SaturValues}
\begin{equation}
g_{xc}\rightarrow \frac{2}{3},
\end{equation}%
\begin{equation}
g_{2x},\,g_{2c}\rightarrow \frac{4}{3},
\end{equation}%
\begin{equation}
G_{xc}\rightarrow \frac{1}{4}.
\end{equation}

\begin{figure}[t]
\includegraphics[scale=0.9]{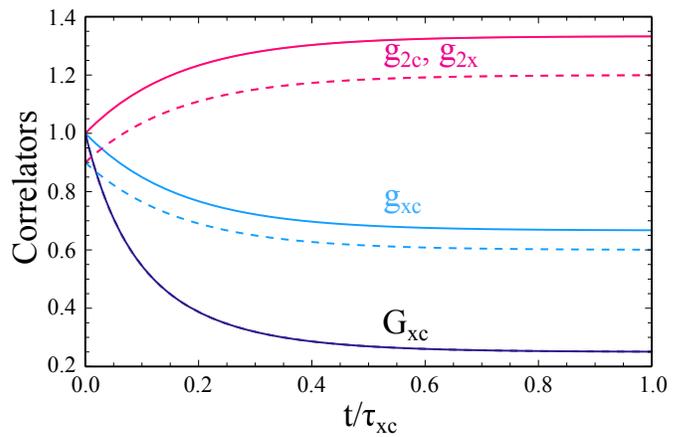}
\caption{(Color online) The time evolution of the correlators $G_{xc}(t)$
(black), $g_{xc}(t)$ (blue) and $g_{2x}(t) = g_{2c}(t)$ (red).
There is initially $N(0)=10$ lower-branch polaritons in the coherent (solid lines) and
and the Fock (dashed lines) states (see text). Infinite lifetimes of excitons and photons
$\tau_{x}=\tau_{c}=\infty$ are assumed.}
\label{fig1}
\end{figure}

These saturations values (\ref{SaturValues}a-c) are quasiclassical and can be obtained
using the equidistribution of particles, i.e., using the density matrix \eqref{GSrepres}
with the kernel $p$ depending on $\mathbf{x}^{2}+\mathbf{y}^{2}$ only. There are quantum
corrections to the expressions (\ref{SaturValues}a-c) where the kernel
$p(\mathbf{x},\mathbf{y})$ is a non-positive and singular function. Consider, e.g., the
evolution of the initial Fock state of $N$ polaritons in the simplest case of infinite
exciton and photon lifetimes. The stochastic mixing process equalizes the probabilities
of all possible combinations of $n_{a}$ and $n_{b}$ such as $n_{a}+n_{b}=N$. Namely,
\end{subequations}
\begin{equation}
P(n_{a},n_{b},t\rightarrow \infty )=\frac{1}{N+1}\delta _{n_{a}+n_{b},N}.
\label{EquiDist}
\end{equation}%
As a result, for $t\rightarrow \infty $ we have $\langle n_{a,b}\rangle =N/2$%
, $\langle n_{a,b}^{2}\rangle =N(2N+1)/6$, $\langle n_{a}n_{b}\rangle
=N(N-1)/6$, so that the factor $(1-N^{-1})$ appears in the right-hand-sides
of the expressions (\ref{SaturValues}a,b).

In Figure \ref{fig2} we demonstrate the behaviour of several diagonal elements
$P(n_{a},n_{b},t)$ of the density matrix as functions of time for the initial coherent
distribution \eqref{PoissonDistr} and realistic exciton and photon lifetimes. In the
panel (a) we account for the finite stochastic exciton-photon conversion time, while in
the panel (b) the stochastic conversion is neglected. In the absence of stochastic
processes there is an asymmetry in the behaviour of the density matrix elements
$P(n_{a},n_{b},t)$ corresponding to the same number of polaritons $N=n_{a}+n_{b}$, which
is caused by the difference of exciton and photon lifetimes. The stochastic process
affects noticeably this behavior, as it is seen in Fig.~\ref{fig2}(a). Due to the
transformations of photons to excitons and vice versa, the density matrix elements
$P(n_{a},n_{b},t)\approx P(n_{b},n_{a},t)$ if $t>\tau_{xc}$. The insets in
Figs.~\ref{fig2}(a,b) show the calculated energy per particle. As expected, this energy
exponentially tends to zero on the timescale of $\tau_{xc}$ [inset in
Fig.~\ref{fig2}(a)], while it is constant in the absence of the stochastic mixing [inset
in Fig.~\ref{fig2}(b)).

\begin{figure}[t]
\includegraphics[scale=0.9]{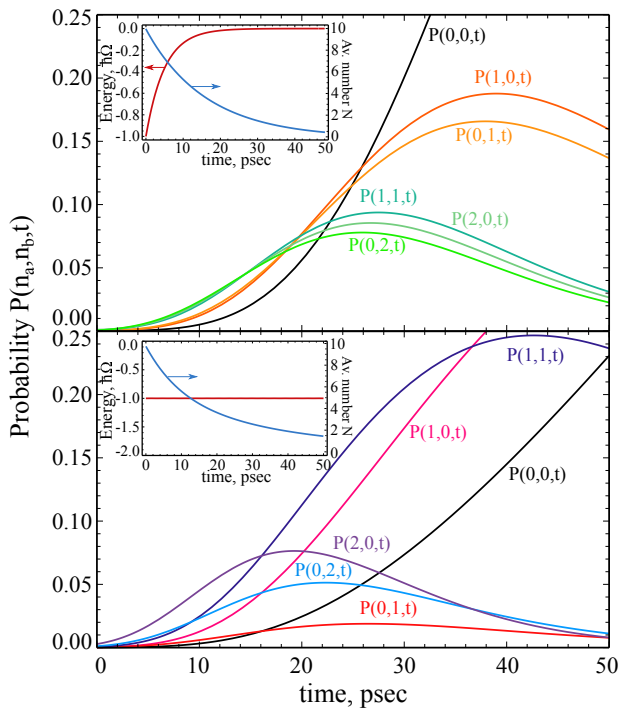}
\caption{(Color online) Showing evolution of some diagonal elements $P(n_a,
n_b, t)$, as well as evolution of the average number of particles $N$ and
the energy per particle on insets. The panel (a) corresponds to the finite
time of the exciton-photon conversion $\protect\tau_{xc}=5\,\mathrm{ps}$,
while the behavior in the absence of stochastic conversion is shown in the
panel (b). The polariton condensate is initially in the coherent state with $N=10$.
The life-times of excitons and photons are $\protect\tau_{x}=40\,\mathrm{ps}$ and
$\protect\tau_{c}=10\,\mathrm{ps}$.}
\label{fig2}
\end{figure}

It is seen that the pronounced variations of the quantum coherence, exciton-photon
correlators and the energy-per-particle with time may be observed if $\tau _{xc}<\tau
_{x},\,\tau _{c}$. How short can be the exciton-photon conversion time $\tau _{xc}$ in
realistic structures? Clearly, in the strong coupling regime one requires $\tau
_{xc}>\Omega ^{-1}$, otherwise the polariton Rabi-oscillations
~\cite{Norris1994,Gibbs,Masha,Dominici2014} would not be observed in time-resolved
optical spectroscopy experiments. Most likely, $\tau _{xc}$ should be of the order of the
decoherence time of Rabi oscillations (typically, 3-10 ps), which can still be short
compared to the polariton lifetime. The most direct way to prove the existence of
stochastic exciton-photon conversions in polariton condensates would be through the
measurement of the generalized exciton-photon correlator (\ref{Gbigxc}). Deviation of
this correlator from 1 would be a signature of the stochastic process described here.
However, the experimental detection of correlations between photocurrent and
photoluminescence is a challenging task. A simpler experiment would consist in
simultaneous measurements of the noise in Kerr rotation angle of a linearly polarized
probe pulse and the noise in the photoluminescence of a polariton condensate excited by a
circularly polarized pump pulse. Indeed, the Kerr rotation angle is proportional to the
exciton fraction of the polariton condensate \cite{Masha}, while the photoluminescence
signal is proportional to its photonic fraction.

Note, that the stochastic exciton-photon conversion might be responsible for the increase
of the second order coherence function $g_{2}(0)$ above the polariton lasing threshold in
the experiments of Kasprzak \textit{et al.} \cite{Richard}. The experimental $g_{2}(0)$
increases even beyond the predicted by our model limit of 4/3. This may be a consequence
of other mechanisms of decoherence, e.g. polariton-polariton scattering and interaction
with bi-excitons.

Finally, while in this Letter we have specifically considered an initial
coherent state of the polariton condensate, which is the most natural choice
for polariton lasers, and only confronted it to the exotic Fock state limit,
our theory can be easily extended to any other statistics of the initial
state (e.g., the thermal statistics). Our goal is to show that the
stochastic conversion of excitons into photons and vice versa should have
strong influence upon the quantum optical properties of exciton-polariton
condensates.

\emph{Acknowledgments.---}We thank N. Gippius and O. Kibis for valuable discussions. This
work was supported in part by EU IRSES projects POLAPHEN and LIMACONA. AVK acknowledges
the financial support from the EPSRC Established Career Fellowship RP008833. ASS
acknowledges fellowship from the Russian President Program and RFBR project 13-02-00944.
IAS acknowledges the support from EU ITN project NOTEDEV and Rannis project BOFEHYSS.


\end{document}